\documentclass[a4paper]{jpconf}
\usepackage{graphicx}
\usepackage{amsmath}
\usepackage{amssymb}

\begin{document}
\title{Time symmetry in wave function collapse models}

\author{Daniel Bedingham}

\address{Faculty of Philosophy, University of Oxford, OX2 6GG, United Kingdom.}

\ead{daniel.bedingham@philosophy.ox.ac.uk}

\begin{abstract}
A framework for wave function collapse models that is symmetric under time reversal is presented. Within this framework there are equivalent pictures of collapsing wave functions evolving in both time directions. The backwards-in-time Born rule can be broken by an initial condition on the Universe resulting in asymmetric behaviour. Similarly the forwards-in-time Born rule can in principle be broken by a final condition on the Universe.

\end{abstract}

\section{Introduction}

\subsection{A time-asymmetric example}
\label{inter}

\begin{figure}[h]
\begin{center}
\includegraphics[width=14pc]{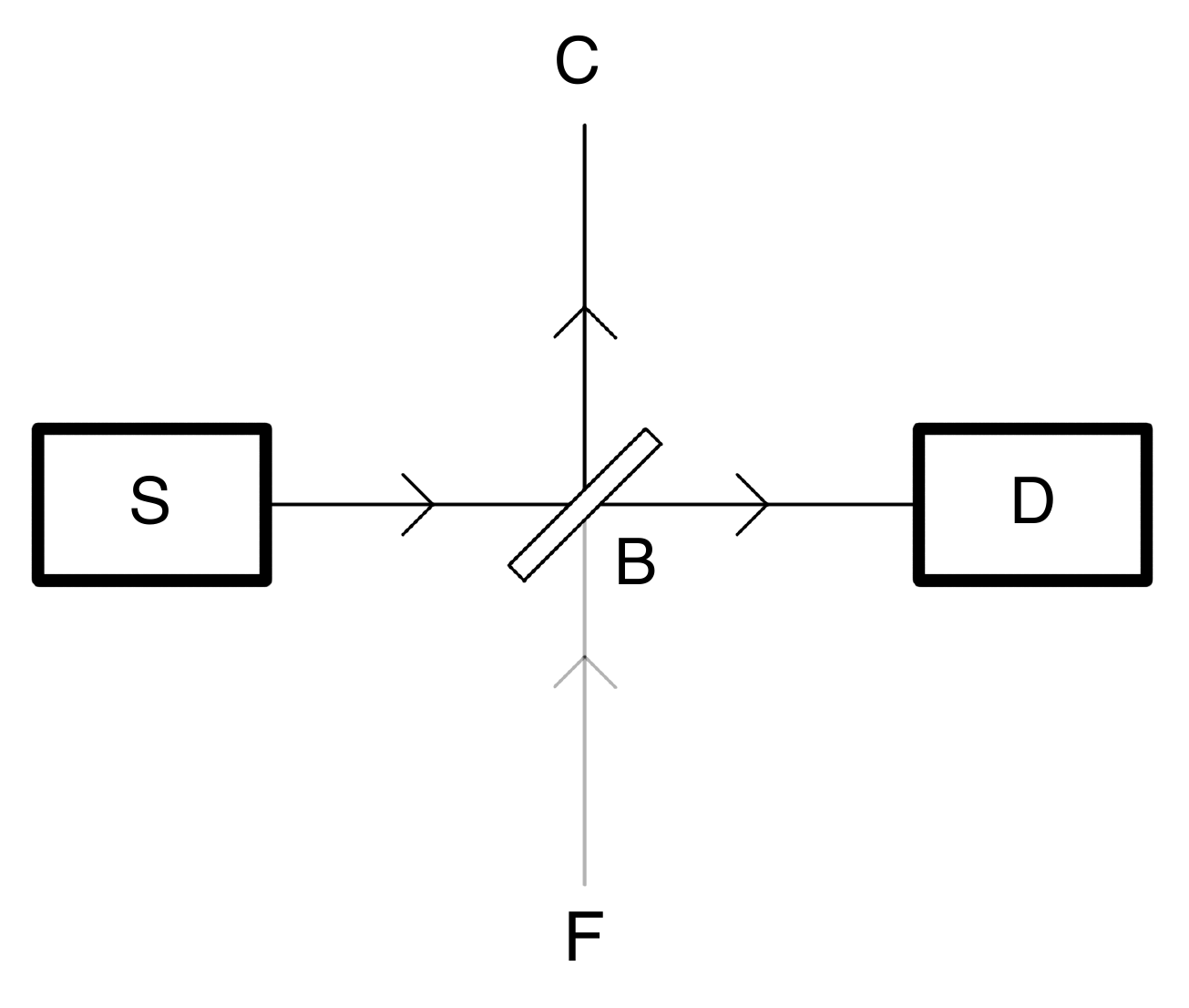}
\caption{Beam splitter thought experiment.}
\label{F1}
\end{center}
\end{figure}

Consider the experimental situation depicted in figure \ref{F1}\footnote{Thanks to Avshalom Elitzur for suggesting this example in a question at this talk.} (taken from reference \cite{pen}). A photon is emitted from a source {\sf S}. It encounters a beam splitter {\sf B} from which it is either transmitted then detected by a detector {\sf D}, or it is reflected up towards the ceiling {\sf C} with no detection made at {\sf D}. The Born rule tells us that a single photon arriving at the beam splitter has a probability of $0.5$ for each possibility. If we perform this experiment many times we can confirm this by comparing the number of detection events with the number of times the source emitted a photon. Both of these types of events are presumably observable to us. The source could be a reliable source of single photons emitting at a fixed rate in which case we do not directly observe individual emission events but we can be fairly sure of the number of emissions in a fixed time period. Perhaps when the detector registers there is a click.

Now consider this experiment backwards in time. That is not to say that we propose to actually go backwards in time or to try to build a source of backwards-in-time photons. We simply consider the events of the experiment which involve a photon interacting either with the source {\sf S}, or with the beam splitter {\sf B}, or with the detector {\sf D}, in the backwards-in-time order. The story starts with a click at {\sf D} as the photon heads backwards in time towards {\sf B} where then with certainty it is transmitted through the beam splitter arriving back at the source (since all photons detected at {\sf D} originated at the source {\sf S}). 

Now if we apply the Born rule to the photon going backwards in time, then as it interacts with the beam splitter there would be a probability amplitude to two possible trajectories. We would predict that only half the photons should arrive back at the source whilst the other half are sent to the floor {\sf F}. In reference \cite{pen} Penrose has argued that this example indicates a time asymmetry in wave function collapse. We might imagine that the backwards-in-time photon, once it encounters the beam splitter enters a superposition state of wave packets at the two locations {\sf S} and {\sf F}. However, the collapse of the wave function, were it to happen in the backwards-in-time direction would always choose the history {\sf SBD}. There are no photons with history {\sf FBD} (let us assume that there is some cold opaque object at {\sf F} to ensure this). The Born rule simply doesn't work for the photon going backwards in time.

In drawing this conclusion we are stripping down the physical system to consider only the behaviour of the photon. It is one of the strengths of quantum mechanics that we can do this. We only need to know the amplitudes for the various options and not all the details and circumstances of other things going on around the experiment. Indeed there will be records in the world that the source is emitting photons at the rate that it is. Perhaps there are other signatures such as a depletion of energy in the source. If the source functions by optical pumping of a small number of atoms then the transparency of the atoms might be dependent on the emission rate. There may be records of the cold opaque object at {\sf F} being kept at absolute zero such that there is simply no possibility for a photon to originate from it. These signatures are part of the extended experiment and they are correlated to the specific behaviour of the photons but the Born rule applied to the photon alone shouldn't care about all this. There is no obligation for the wave function of the photon to collapse to those particular outcomes that result in correlations with the wider world.

\subsection{Inherent time asymmetry or the effect of an initial condition?}

There seems to be no doubt that the Born rule doesn't work in the reverse-time direction in the example of figure \ref{F1}. There is then the possibility that the Born rule has an inherent time asymmetry---that it only works in the forwards-in-time direction. But there is also the possibility that the time asymmetry is due to an initial boundary condition which serves as a constraint. Furthermore, this initial constraint could be the (low-entropy) initial condition of the Universe. This view is in line with the standard explanation of time asymmetric behaviour in the world. For the photon travelling backwards in time, having its wave packet split into a superposition by the beam splitter, it is not simply left with a $50$-$50$ choice of collapse to {\sf F} or collapse to {\sf S}, it must also weigh this up against the need of the Universe to achieve its low entropy initial state. By collapsing the wave packet at {\sf S} many correlations are established with the extended experiment and entropy is reduced.

We might also consider the possibility of some far-off final condition of the Universe which would ultimately cause a breakdown of forwards-in-time Born-rule behaviour. Note that in this picture the boundary conditions are responsible for modifying the probabilistic behaviour of the world in order to fulfil the condition. This is characteristically different from a deterministic world where the state at any given moment must be such that the boundary conditions are satisfied when we evolve in time.

It is therefore by no means certain that collapse of the wave function is inherently time asymmetric. However, if we are to treat collapse as a genuine physical process we face further issues to do with time symmetry.

%%%%%%%%%%%%%%%%%%%%%%%%%%%%%%%%%%%%%%%%%%%%%%%
\section{Collapse models}

In models of wave function collapse we regard the wave function as physically real and the collapse of the wave function as a real physical process. This requires a deviation from the standard Schr\"odinger dynamics. The general idea is to modify the Schr\"odinger equation in such a way that certain types of large-scale superposition states are unstable and will collapse. Yet the modified equation should be universally applicable and so when applied to micro systems it should be well approximated by the standard Schr\"odinger equation. These modifications need to be both stochastic (since collapse is random) and non linear (since the chance of a collapse outcome depends on the state itself). The most well known of these are the GRW model \cite{GRW} and the CSL model \cite{CSL1,CSL2}. There are also relativistic versions of these models \cite{REL1,REL2,REL3}.

\subsection{The GRW model}
\label{GRW}

Let us briefly summarise the GRW model for the sake of a definite example. The GRW model concerns a set of $N$ distinguishable particles with wave function $\psi_t(x_1,\ldots,x_2)$. The wave function usually satisfies the usual Schr\"odinger equation but at random times chosen with fixed probability per unit time per particle $i$, the wave function makes a jump of the form
\begin{equation}
\psi_t\rightarrow \psi_{t+} = j(z-x_i)\psi_t,
\label{jump}
\end{equation}
with $z$ a random variable. This happens for every particle $i$ independently. Each particle has its own random sequence of jumps occurring at a different set of random times. The jump operator $j$ takes the form
\begin{equation}
j(x) = \frac{1}{\left(\pi a^2\right)^{1/4}}\exp\left(\frac{-x^2}{2a^2}\right),
\end{equation}
where $a$ is a fixed parameter with units of length chosen by GRW to be $10^{-7}{\rm m}$. The action of the jump operator is that of a quasi projection of the $i$th particle about position $z$. The complete set of operators $\{j^2(z-x_i)  |z \in \mathbb{R} \}$ form a POVM. All the collapses occur in this smeared position state basis. This is the preferred basis of this model. It is a remarkable feature that even with collapses only occurring in the position basis, models of this type are able to account for collapse of the wave function in any conceived measurement situation. 

Finally, the random collapse centres $z$ are drawn from a probability distribution of the form
\begin{equation}
\mathbb{P}_t\left( z  \right) = \frac{\int dx_1\cdots dx_N|j(z-x_i)\psi_t|^2}{\int dx_1\cdots dx_N|\psi_t|^2}.
\label{prob}
\end{equation}
This is nothing but the Born rule probability distribution for a quasi projection of the form (\ref{jump}). This makes the model non linear. 

The rate of collapse per particle is chosen to be sufficiently small that individual particles very rarely undergo jumps (GRW choose a rate of $10^{-16}{\rm s}^{-1}$, this is about once every lifetime of the Universe). The physics of a few particles is therefore left unaffected. However, for bulk matter with of order $10^{24}$ particles, there are many jumps each second causing, for example, pointers to commit to definite readings etc. The theory does the job of making a very minor modification to the micro physics of individual particles which leads to radical effects at the macro level. Similar principles are used in other collapse models.

\subsection{Time reversal symmetry in collapse models}
Now let us return to the question of time reversal symmetry. When a single jump occurs, the wave function of a single particle goes from being dispersed in space to suddenly being localised about the random position $z$. This is clearly a time-asymmetric process. Viewed backwards in time what we have is a localised wave function spontaneously becoming dispersed in space. When the jump happens it affects only the future state of the particle and this introduces a preferred time direction.

Note that often solutions to the Schr\"odinger equation seem to display time directed behaviour such as quantum dispersion. However, it is straightforward to show the that reverse-time process, quantum focussing let's call it, is also a solution to the Schr\"odinger equation\footnote{provided that we remember to take the complex conjugate of the wave function}. Quantum dispersion is more common than quantum focussing simply because the initial conditions for quantum focussing are not prevalent in the Universe. 

But the situation with the GRW jumps is different. We can't make the time reverse of a jump look like a jump. The dynamical rule which is composed of the Schr\"odinger equation with jumps does not apply to a reverse time sequence of wave functions. Faced with a sequence of wave functions and asked to identify the forward direction of time, it can be done if a jump occurs (but not if one doesn't).

One way to restore time reversal symmetry to this picture is to introduce a separate backwards-in-time wave function (we can call it $\bar{\psi}$). This new wave function might look quite similar in form to $\psi$ but it has the jump rule
\begin{equation}
\bar{\psi}_t\rightarrow \bar{\psi}_{t-} = j(z-x_i)\bar{\psi}_t.
\label{jumprev}
\end{equation}
The jumps affect the past state of $\bar{\psi}$ rather than the future state. This is the same jump rule as for $\psi$ but acting in the reverse-time direction. To be clear, the jump rule is equivalent to a pair of left and right boundary conditions on $\bar{\psi}$ at time $t$: on the $t+$ side the state is $\bar{\psi}_t$; on the $t-$ side the state is $ j(z-x_i)\bar{\psi}_t$. At all other times the wave function satisfies the (time-reversal-symmetric) Schr\"odinger equation.

What ties the two wave functions $\psi$ and $\bar{\psi}$ together? At a given moment they might look broadly similar and make almost identical predictions. However, what guarantee is there that this will be the case at other times? And if they are not the same which one should we trust to tell us about the state of the world?

We can in fact ensure that the two wave functions are empirically equivalent at all times by having their jumps occur at identical locations in space and time. To understand this consider a single particle which suffers a sequence of jumps at times $t_1<t_2<\cdots<t_{n-1}<t_n$ and at locations $z_1,z_2,\ldots,z_{n-1},z_n$. The forwards-in-time wave function $\psi$ undergoes first a jump about $z_1$ at $t_1$, then a jump about $z_2$ at $t_2$, etc. By contrast the evolution of the backwards-in-time wave function $\bar{\psi}$ involves first a jump about $z_n$ at time $t_n$, then a jump about $z_{n-1}$ at time $t_{n-1}$, etc. The history of the particle is the same in each case as it passes through the same series of locations at the same times. This is despite the fact that the detailed behaviour of the two wave functions will be quite different from one another. In particular the cycle of sudden collapse followed by dispersion will be occurring in opposite time directions.

The same idea works when we have more particles. We have consistency between $\psi$ and $\bar{\psi}$ if they share the same set of collapse centres $\{z\}$. It has even been argued that the ontology of collapse models should be just this set of collapse centres \cite{BELL}. As Bell puts it, the collapse centres {\it ``...are the mathematical counterparts in the theory to real events at definite places and times in the real world. [] A piece of matter then is a galaxy of such events."}\cite{BELL}. The collapse centres are the part of the theory that exist in ordinary space and time (the wave function exists is an infinite dimensional Hilbert space). It therefore makes sense to treat the collapse centres as the basis for local beables in the theory.

Further developing this picture it has been shown in reference \cite{HERB2} for a lattice model of wave function collapse that, given a set of collapses generated by some specific wave function, a generic wave function (subject to some constraints such as consistent particle number) will, after undergoing this same set of collapses, tend towards the original wave function. This implies that the information contained in the wave function at any given point is also contained in the collapse data from a sufficient period of history. We might therefore be able to abandon the wave function altogether \cite{KENT}. Of course these considerations will apply to both the forwards and backwards in time wave functions.

In order to demonstrate that this picture is indeed time symmetric it must be the case that the probability rule (\ref{prob}) is also true for $\bar{\psi}$, i.e.
\begin{equation}
\bar{\mathbb{P}}_t\left( z  \right) = \frac{\int dx_1\cdots dx_N|j(z-x_i)\bar{\psi}_t|^2}{\int dx_1\cdots dx_N|\bar{\psi}_t|^2}.
\label{probrev}
\end{equation}
This has been shown in reference \cite{me} for two different models of wave function collapse. (Specifically these models are the lattice model of collapse of references \cite{HERB2,LATT1,HERB1} and the QMUPL model of reference \cite{DIOS} in the localised particle limit.) The demonstration involves a statistical test. A forwards-in-time wave function is used to generate a set of collapses according to rules of the type outlined in section \ref{GRW}. A test wave function is then used as a backwards-in-time wave function and evolved backwards using (\ref{jumprev}) with the same set of collapses in the reverse order. The likelihood that these collapses could have instead been generated by the backwards-in-time probability rule (\ref{probrev}) is then tested by calculating a statistic designed to indicate this (along the lines of the Kolmogorov-Smirnov test).

It is found that to within statistical error the probability rule (\ref{probrev}) (which is essentially the Born rule) works perfectly well backwards in time. What this means is that if you were given the set of collapses $\{z\}$ along with the two wave functions $\psi$ and $\bar{\psi}$ (of course not being told which was which), there would be no way for you to determine by analysing all this information, which was the forward direction of time and which was the backward direction of time. The dynamical collapse rules work identically in both directions. The collapse model is time reversal symmetric.

\subsection{The effect of an initial condition}
As we have seen in section \ref{inter}, this conclusion cannot be universally valid. We can envisage examples such as the beam splitter experiment of figure \ref{F1} where the Born rule is not expected to hold backwards in time. However, we have argued that this does not necessarily imply an inherent time asymmetry in the collapse of the wave function. A possible explanation is that it is the effect of a low entropy initial condition. Indeed neither of the examples in reference \cite{me} involved large numbers of particles with highly ordered initial conditions. Moreover, the precise details of the initial wave function are explicitly washed away after a sufficient period of time in the lattice model of reference \cite{HERB2}. It is conceivable that in these cases the effect of the initial condition is innocuous or that the system is in a state of equilibrium. The idea that asymmetries result from boundary conditions at least allows for an explanation for the observed time symmetry in these examples which would be harder to do otherwise.

We can propose a modification of the backwards-in-time probability rule valid in the presence of a fixed initial condition on the state $\bar{\psi}_{0}$,
\begin{equation}
\bar{\mathbb{P}}_t\left( z  \right) \rightarrow \bar{\mathbb{P}}_t\left( z  |  \bar{\psi}_{0} \right)
 = \frac{\bar{\mathbb{P}}_t\left( z  \cap \bar{\psi}_{0} \right)}{\bar{\mathbb{P}}_t\left( \bar{\psi}_{0} \right)}.
\label{probrevmod}
\end{equation}
Here $\bar{\mathbb{P}}_t\left( \bar{\psi}_{0} \right)$ means the probability that the state $\bar{\psi}$ will take the form $\bar{\psi}_{0} $ at the start of the Universe as a result of the backwards-in-time collapse dynamics, given that it takes the form $\bar{\psi}_{t}$ at time $t$. Typically this would be very complicated to determine based on all possible collapse events happening between time $t$ and the start of the Universe.

Similarly a final time boundary condition on the state $\psi_{f}$ would modify the forwards-in-time probability rule
\begin{equation}
\mathbb{P}_t\left( z  \right) \rightarrow \mathbb{P}_t\left( z  | \psi_{f} \right)
 = \frac{\mathbb{P}_t\left( z  \cap \psi_{f} \right)}{\mathbb{P}_t\left(\psi_{f}\right)}.
\end{equation}
These modified probability rules potentially account for the time asymmetric behaviour of the Born rule identified in the Introduction in a way which retains the time symmetry of the collapse dynamics. In terms of the beam splitter example, the probability rule (\ref{probrevmod}) would weight the probabilities in favour of the backwards-in-time photon being transmitted rather than reflected. (Of course the photon itself would not be expected to undergo a GRW collapse, the collapse would happen when the photon has become entangled with the displacement of a sufficient number of atoms in one of the devices of the experiment.)

We finally note that collapse models are typically known to result in a gradual increase in energy. If this happens to the future it cannot also happen to the past. But again a simple reason for this behaviour is that energies in the past are constrained by the initial condition.

\section{Summary}
We have argued that wave function collapse models can be understood in a way in which the dynamical rules are symmetric under time reversal symmetry. This implies that the physical process of the collapse of the wave function is symmetric in time. We have argued that certain examples where the Born rule does not hold in the reverse time direction can be attributed to the effect of an initial condition of the Universe. This is the standard explanation of time asymmetric behaviour resulting from time symmetric laws.

%\subsection*{Acknowledgments}
\ack
This work was supported by a grant from the Templeton World Charity Foundation.


\section*{References}
\begin{thebibliography}{9}
\bibitem{pen} Penrose R 2004 {\it The Road to Reality} (London: Jonathan Cape) section 30.3 pp 819-823
\bibitem{GRW} Ghirardi GC, Rimini A and Weber T (1986) Phys.~Rev.~D {\bf 34} 470
\bibitem{CSL1} Pearle P 1989 Phys.~Rev.~A {\bf 39} 2277
\bibitem{CSL2} Ghirardi GC, Pearle P and Rimini A 1990 Phys.~Rev.~A {\bf 42} 78
\bibitem{REL1} Tumulka R 2006 J.~Stat.~Phys.~{\bf 125} 821
\bibitem{REL2} Bedingham D 2011 Found.~Phys.~{\bf 41} 686
\bibitem{REL3} Bedingham D 2011 J.~Phys.: Conf.~Ser.~{\bf 306} 012034
\bibitem{BELL} Bell J 2004 {\it Speakable and Unspeakable in Quantum Mechanics} (Cambridge) chapter 22 pp 201-212
\bibitem{HERB2} Dowker F and Herbauts I 2005 Found.~Phys.~Lett.~{\bf 18} 499
\bibitem{KENT} Kent A 1989 Mod.~Phys.~Lett.~A {\bf 4} 1839
\bibitem{me} Bedingham D and Maroney O 2015 {Time reversal symmetry and collapse models} {\it Preprint} quant-ph/1502.06830
\bibitem{LATT1} Dowker F and Henson J 2004 J.~Stat.~Phys.~{\bf 115} 1349
\bibitem{HERB1} Dowker F and Herbauts I 2004 Class.~Quant.~Grav.~{\bf 21} 2963
\bibitem{DIOS} Di\'osi L 1988 Phys.~Lett.~A {\bf 132} 233
\end{thebibliography}
\end{document}